\documentclass[runningheads]{llncs}
\usepackage{graphicx}
\usepackage{cite}
\usepackage{amsmath,amssymb,amsfonts}
\usepackage{graphicx}
\usepackage{textcomp}
\usepackage{xcolor}

\begin{document}
\title{\textit{RewardRating}: A Mechanism Design Approach to Improve Rating Systems}
%
%
\author{Iman Vakilinia\inst{1} \and
Peyman Faizian\inst{2} \and
Mohammad Mahdi Khalili\inst{3}}
\authorrunning{I. Vakilinia et al.}
%

\institute{University of North Florida, FL, USA \\
\email{i.vakilinia@unf.edu} 
\and
Florida State University, FL, USA\\
\email{faizian@cs.fsu.edu} 
\and
University of Delaware, DE, USA\\
\email{khalili@udel.edu}
}
\maketitle              
\begin{abstract}
Nowadays, rating systems play a crucial role in the attraction of customers for different services. However, as it is difficult to detect a fake rating, attackers can potentially impact the rating's aggregated score unfairly. This malicious behavior can negatively affect users and businesses.
To overcome this problem, we take a mechanism-design approach to increase the cost of fake ratings while providing incentives for honest ratings. Our proposed mechanism \textit{RewardRating} is inspired by the stock market model in which users can invest in their ratings for services and receive a reward based on future ratings.
First, we formally model the problem and discuss budget-balanced and incentive-compatibility specifications. Then, we suggest a profit-sharing scheme to cover the rating system's requirements. Finally, we analyze the performance of our proposed mechanism. 

\end{abstract}

\section{Introduction}

\begin{figure}[ht]
	\begin{center}
		\includegraphics[width=4in,height=2in,keepaspectratio]{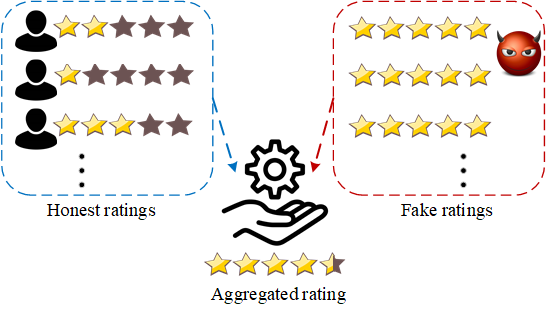} 
		\caption{An example of sybil attack to rating system}
		\label{fig:syb}
	\end{center}
	\vspace{-.2in}
\end{figure}

Recently, online rating systems have become a significant part of potential customers' decisions. 
According to a survey~\cite{brightlo}, 90\% of consumers used the Internet to find a local business in 2019, businesses without 5-star ratings risk losing 12\% of their customers, and only 53\% of people would consider using a business with less than 4-star ratings. 
Due to the importance of such ratings, malicious users attempt to impact the rating of a service\footnote{A rating system can be applied to different entities such as a service, product, community, or a business. For the rest of the paper, we refer to such an entity as a service.} by submitting fake scores. For example, a malicious service owner would submit fake 5-star ratings to increase his service's aggregated rating, as depicted in Figure~\ref{fig:syb}. On the other hand, a malicious competitor would submit fake low ratings to subvert rivals' reputation. 
Moreover, rating systems are vulnerable to \textit{sybil} attacks in which a malicious entity can forge fake users and utilize them for fake ratings. Such a vulnerability caused the advent of companies offering on-demand fake reviews and ratings~\cite{nyt}.

Considering the lack of an appropriate safeguard to prevent fake ratings, in this paper we take a novel mechanism-design approach to increase the malicious users' cost for submitting fake ratings while providing incentives for honest raters. Our proposed mechanism \textit{RewardRating}, is inspired by the stock market in which reviewers can invest in their ratings for services and receive a reward based on their investments. \textit{RewardRating} is equipped with a profit-sharing scheme to satisfy incentive-compatibility and budget-balanced properties as we discuss in the next sections.
The main contributions of this paper are the two parts, as described below:\\
(1) We propose a new mechanism to disincentivize fake ratings while stimulating honest ratings. \\
(2) We investigate a design of an incentive-compatible self-sustained profit sharing model to be applied for a rating system.

The rest of the paper is organized as follows. In the next section, we review related works. The system model is presented in section 3. Details of our proposed mechanism is described in section 4. We analyze our proposed mechanism in section 5. Finally, we conclude the paper in section 6.

\section{Related Work}

To support consumers and services, the US Federal Trade Commission (FTC) takes legal actions against fake reviewers~\cite{FTC}. Furthermore, rating platform providers such as Amazon, Google, and Yelp have restricted policies for fake reviews and ban the incentivized reviews in which service owner provides incentives in return for positive reviews~\cite{amazon,google,yelp}. 
Different safeguards can be implemented to reduce the number of fake reviews. For instance, a review platform can verify the identity of a reviewer before allowing the review submission, and a reviewer should have a valid email-address and phone-number. Furthermore, review platforms can use machine-learning tools to detect and remove fake reviews~\cite{yelpfilter}. 
Although filters for checking the authenticity of reviews are necessary, studies show that still many reviews and ratings are fake, and filters cannot prevent them~\cite{fox,hustle}. 
On the other hand, despite the fact that \textit{sybil} attacks have been widely studied in various networks~\cite{kumar2020game, levine2006survey, yu2011sybil}, such studies especially targeted towards preventing \textit{sybil} attacks on rating systems are few and far between.

Detection of fake reviews using machine-learning techniques have been studied widely in the literature~\cite{hajek2020fake,ahmed2018detecting,heydari2015detection,wu2020fake,yao2017automated}. These methods use linguistic features of a review, meta-data associated with a review, and the service history to check the validity of reviews. 
However, the task of detecting fake ratings is more challenging, mainly due to the fact that users might have different preferences or expectations for a service. For instance, two authentic and independent users might receive the same service and truthfully rate it quite differently (1-star vs. 5-star). It is difficult to detect if one of these ratings is fake, or they are just simply referring to different aspects of a service (e.g., quality of food vs. air-conditioning in a restaurant). 
On the other hand, Monaro~\textit{et al.}~\cite{monaro2020spotting} studied the detection of fake 5-star ratings using mouse movements. This study discovered users spend more time and wider mouse trajectories to submit false ratings. 

In contrast with previous works, in this study, we propose a novel mechanism design approach to improve the quality of the aggregated ratings for the services by increasing the cost for fake ratings while incentivizing honest ratings. Note that \textit{our proposed mechanism's goal is not to detect fake ratings but to incentivize honest rating}s, as discussed in the following.

\section{System Model}

Let $R = \{r_1 , ... , r_n  \}$ be the strictly totally ordered set which represents rating scores that reviewers can assign to a service. We indicate a rating $r_j$ is higher than $r_i$ ($r_i < r_j$) if and only if $i < j$. For example, in google review system, $n$ is 5, and $r_5$ indicates a 5-star rating which is a higher rating compared to $r_2$ which represents a 2-star rating. For the sake of simplicity and without loss of generality, we consider $n=5$ for the examples we present in the rest of the paper. 
Let $\mathcal{U} = \{u_1 , u_2, ... \}$ represent the set of reviewers who submit ratings for a service. 
There is no limitation on the number of reviewers. 
Moreover, a reviewer can submit multiple ratings under different identifiers for the same service. In other words, the system is vulnerable to the \textit{sybil} attack. For example, a service owner can submit unlimited 5-star ratings for himself, or a competitor can submit unlimited 1-star ratings for a rival service. This can be achieved by recruiting fake reviewers.


Reviewers submit ratings to obtain their desired outcomes. We assume there are three types of reviewers based on their intents:
\begin{itemize}
	\item \textbf{\textit{Attacker}}- Submits false ratings to change the aggregated rating score to his preference. For example, a restaurant owner wants to increase his restaurant's rating score by submitting fake 5-star ratings. 
	
	\item \textbf{\textit{Honest}}- Submits the rating truthfully based on his evaluation of a service's quality. 

	\item \textbf{\textit{Strategic}}- Submits the rating to increase his payoff from the system. 
	
\end{itemize}

The main objective of the mechanism is to decrease the number of fake ratings by increasing the cost for \textit{attackers}. On the other hand, we want to provide incentives for \textit{honest} reviewers and motivate \textit{strategic} players to invest in an honest rating. We aim to achieve these goals by making a market where reviewers invest in their ratings, and the rating system rewards the reviewers based on future investments.

\section{Marketizing ratings}

The main idea of \textit{RewardRating} is to create a market for ratings where reviewers invest in their ratings. 
This idea is inspired by the stock market. In the stock market, investors invest in a business based on their prediction of a business's performance in the future when they want to sell their stocks. 

Mapping this to the rating system, in \textit{RewardRating}, reviewers invest in a service's performance. To make it more clear, let us continue with an example:

\textit{Assume Alice goes to a restaurant, and she is happy about the restaurant's service. Alice thinks this restaurant deserves a 5-star rating. However, the current aggregated rating for this restaurant is 3-star. Using \textit{RewardRating} mechanism, Alice can invest in a 5-star score for the restaurant's service. If future reviewers agree with Alice and rate the restaurant higher than 3-star, then Alice has made a successful investment, and as a result, she receives a reward from the system. }

\subsection{Requirements}

\textit{RewardRating} should satisfy the following requirements:

\begin{itemize}
	\item \textbf{\textit{Budget-Balanced}}:
	The system should be self-sustained, as there are no external financial subsidies. In other words, the total asset in the system should be supported by the reviewers. 
	\item \textbf{\textit{Incentive-Compatibility}}:
	The system should provide incentives for truthful reviewers. On the other hand, the system should increase the cost for attackers.
\end{itemize}
\subsection{Mechanism Narrative}

Specifying requirements, now let's study the design of a mechanism that satisfies these properties. The design objective is to place a set of rules for the rating system's game to meet the aforementioned requirements.
A mechanism can be specified by a game $g: \mathcal{M} \rightarrow \mathcal{X} $ where $\mathcal{M}$ is the set of possible input messages and $\mathcal{X}$ is the set of possible outputs of the mechanism.
In the rating system model, players are reviewers. A player chooses his strategy to increase his utility. 
A player's strategy (i.e., the mechanism's input), is the rating, corresponding investment, and the time of investment. 
Finally, the mechanism's output is the aggregated rating of the service and the profit, which is shared among stakeholders.

In \textit{RewardRating}, each rating is accompanied by a corresponding stock value. Reviewers can invest in a rating for a service, by buying the stock associated with it. Let us use the term \textit{coin} to represent the smallest unit for such rating stocks. A new coin is minted for a rating's stock when a reviewer requests to buy such a coin from the rating system.\footnote{Note that, here, coins are \textit{virtual} assets, and minting a new coin means that the system adds to the total value of a rating's stock.} The price of buying a coin from the system is fixed.
Users can sell their coins to the rating system. The selling price of a coin is also fixed; however, the price of buying a coin from the system is higher than the price of selling a coin to the system. 
The difference in the buying price and selling price is a fund that is shared among stakeholders as profit.

\begin{figure*}[ht]
	\begin{center}
		\includegraphics[width=5in,height=3in,keepaspectratio]{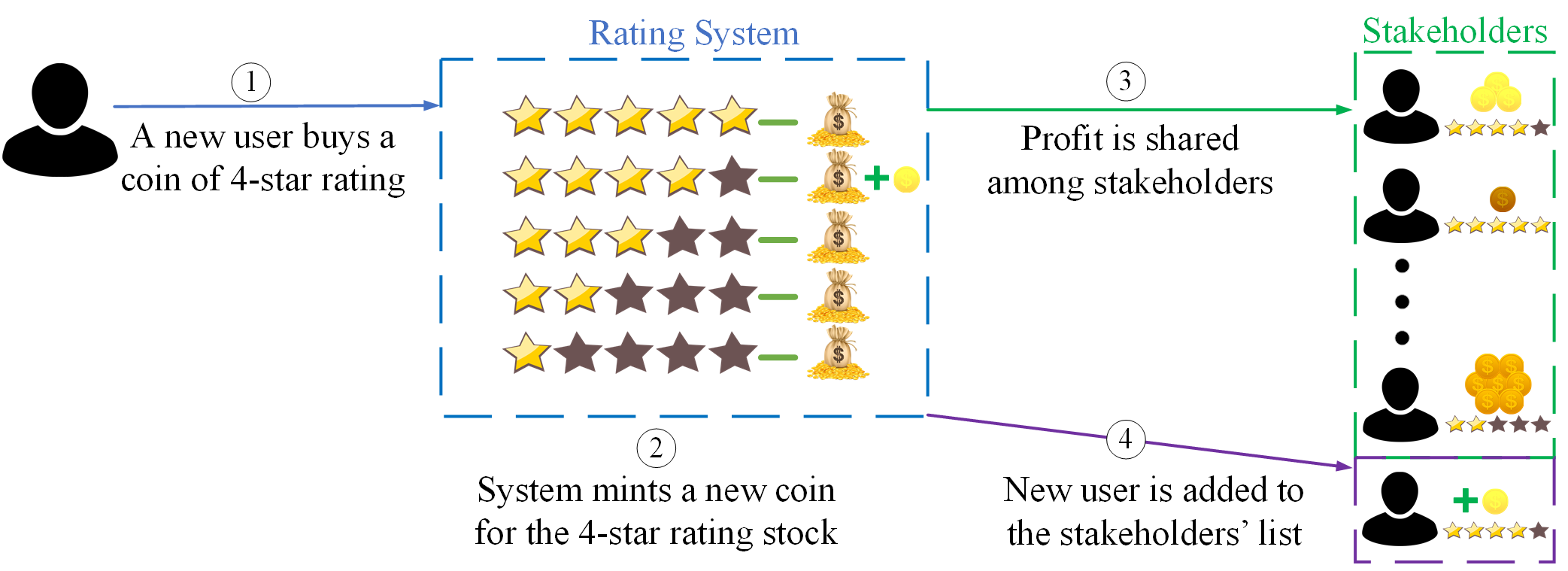} 
		\caption{Buying a coin from the rating system. We have used different colors to demonstrate the difference in coins of ratings' stocks.}
		\label{fig:buy}
	\end{center}
	
\end{figure*}

\begin{figure*}[ht]
	\begin{center}
		\includegraphics[width=5in,height=3in,keepaspectratio]{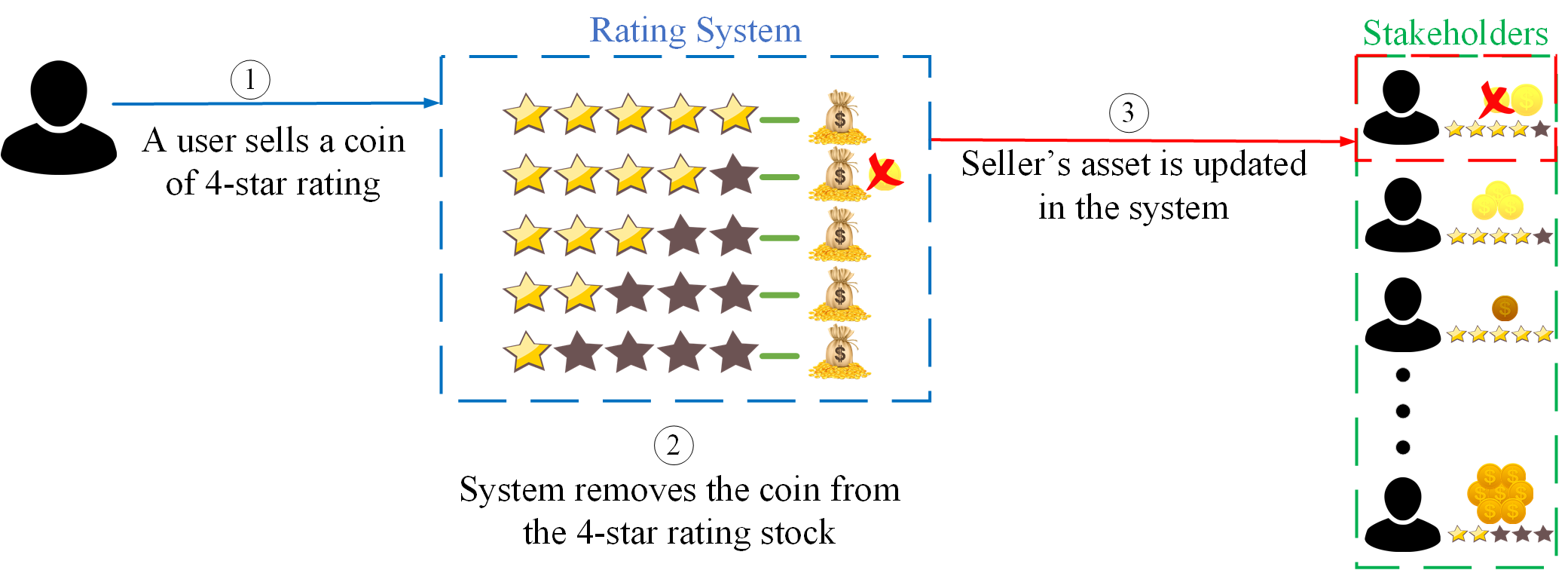} 
		\caption{Selling a coin to the rating system. We have used different colors to demonstrate the difference in coins of ratings' stocks.}
		\label{fig:sell}
	\end{center}
\end{figure*}

We discuss the details and the reasoning behind the design decisions made, in the next sections. 
Figures~\ref{fig:buy} and~\ref{fig:sell} demonstrate the overall picture of buying/selling coins from/to the rating system. 

\subsection{Mechanism Components}

In this section, we formally define the components of \textit{RewardRating} mechanism. 
Let $C = \{c_1 , ... , c_n  \}$ represent the set of $n$ types of coins for ratings available in the rating system such that $c_j$ represents coins for the score $r_j \in R$.
Let $\alpha \in \mathbb{R}^+$ represent the price of buying a coin from the system. 
Let $x_{i,j}^t \in \mathbb{R}^+ $ represent the number of $c_j$ coins that user $u_i$ owns in the rating system at time $t$.
Let $S_t=<x_{i,j}^t, ... , x_{k,l}^t >$ represent the stakeholders in the rating system at time $t$. 
Once a user $u_i$ pays $\alpha$ to the rating system to buy a new coin $c_j \in C$, the system mints a new coin, and updates the stakeholders list accordingly. 

On the other hand, users can sell their coins to the rating system. Let $\beta \in \mathbb{R}^+$ represent the price that the rating system pays to a user in return for a coin. Then, we have $\alpha = \beta + \gamma$ in \textit{RewardRating}. Here, $\gamma \in \mathbb{R}^+$ is a profit that the system earns from selling a new coin. Such a profit is distributed among stakeholders as a reward. Once a user sells a coin to the system, the system removes that coin from the corresponding rating stock and updates the stakeholders list.

For example, assume $u_i$ buys a new coin of 4-star rating with the price of $\$1$ (i.e., $\alpha=1$). Assume $u_i$ decides to sell his coin to the rating system later on, and the price of selling a coin is $\$0.9$ (i.e.,  $\beta = 0.9$). Then, $u_i$ receives $\$0.9$ from the rating system. In this example, the system earns $\$0.1$ profit (i.e.,  $\gamma=0.1$), which is shared among stakeholders.
Note that only the profit of the first minted coin of a service is earned by the rating system's owner. This is due to the fact that for the first coin, there is no previous stakeholder to share the profit.

Let $|c_j^t|$ represent the number of $c_j$ coins minted in the rating system at time $t$. 
Let $\sigma^t $ be the aggregated score of a service at time $t$. The system calculates $\sigma^t$ based on the total investments in the rating stocks for a given service as follows:
\begin{equation}
\sigma^t  =  \frac{ \sum_{\forall c_j \in C} ( |c_j^t|  \times  j) }{ \sum_{\forall c_j \in C} (|c_j^t|) }
\end{equation}

In other words, the aggregated score of a service is calculated based on the total investments on ratings' stocks.
\begin{figure*}[h]
	\begin{center}
		\includegraphics[width=4.9in,height=2.9in,keepaspectratio]{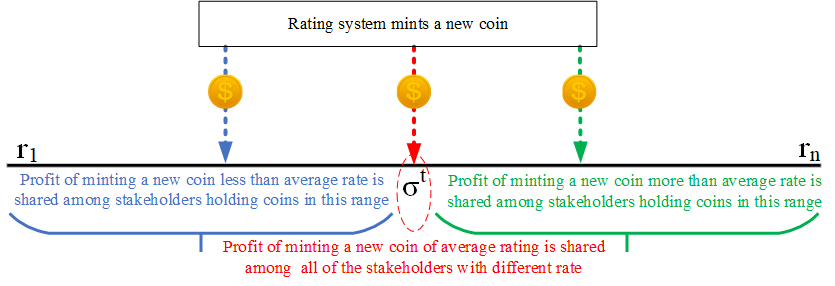} 
		\caption{Sharing profit among stakeholders}
		\label{fig:prof}
	\end{center}
	\vspace{-.15in}
\end{figure*}

\vspace{.1in}
\textit{\textbf{Profit Sharing}}

\textit{RewardRating} collects a profit from minting every new coin. This profit is the difference in the price of buying a new coin by a user from the rating system and the price of selling that coin to the rating system which can be calculated as $\gamma = \alpha - \beta$.  
The mechanism distributes such a profit among stakeholders strategically to satisfy the mechanism requirements. The main challenge here is to minimize the profit that attackers can earn from the system. As if attackers earn profit from the system, then \textit{RewardRating} encourages attackers instead of honest reviewers. 
This is a big challenge as it is hard to distinguish between honest users and attackers in the system.  
To solve this problem, we consider the fact that if a service's rating is affected by fake ratings, then the aggregated rating is biased toward one direction. If attackers submitted fake 5-star ratings, then the aggregated rating is biased toward 5-star, and if attackers submitted fake 1-star ratings, then the aggregated score is biased toward 1-star rating. Considering this fact, the system shares the profit of a new minted coin as follows:

\begin{itemize}
	\item If the new minted coin is for rates higher than the aggregated score, this will in turn increase the aggregated score. In this case, the profit of the new minted coin is shared among those stakeholders holding coins of higher rates than the current aggregated rating. 
	
	\item If the new coin is for rates lower than the aggregated score, this will in turn decrease the aggregated score. In this case, the profit of the new minted coin is shared among those stakeholders holding coins of lower rates than the current aggregated rating. 
	
	\item If the new coin matches the aggregated rating, then the aggregated score does not change, and the profit is shared among all of the rating system's stakeholders. Note that this case is rare, as the coin types (i.e., the options for ratings) does not necessarily match the number of possible values for the aggregated score. For example, in the google review system, we assume users have 5 options for selecting rates (1-star,..., 5-star); however, the aggregated rating has 1 decimal point, which makes 50 possible options for the aggregated score. 
\end{itemize}

To model such a profit sharing, first, we need to define a set of stakeholders who earn profit from the system once there is a new investment. Let $c_j$ represent the type of a new minted coin at time $t$. Let $\mathcal{W}_j^t \subseteq \{1,2,...,n\}$ represents the set of indices of ratings' stocks which their stakeholders are rewarded for the new minted coin $c_j$. In other words, a user does not receive a reward if he does not own a rating coin in the set of $\mathcal{W}_j^t$ when a new coin $c_j$ is minted at time $t$.
Then, we model $\mathcal{W}_j^t$ as:
\begin{equation} 
\mathcal{W}_j^t =
\begin{cases} 
\{i \in \mathbb{N} : \sigma^t < i \leq n \}  & j > \sigma^t  \\
\{i \in \mathbb{N} : 1 \leq i < \sigma^t \}  &  j < \sigma^t    \\
\{i \in \mathbb{N} : 1 \leq i \leq n\}  & j = \sigma^t 
\end{cases}
\label{eq:win}
\end{equation}

\textit{RewardRating} distributes the profit in a way that stakeholders are earning more profit if they have the rating coins closer to the rating of the new minted coin. In other words, with the growth of distance between $j$ and $w \in \mathcal{W}_j^t$, the profit of a stakeholder of a coin $c_w$ is decreasing. 
To model this, let $f(w,j) \in \mathbb{R}^+ $ represent a function such that $\frac{ \partial f(w,j) }{ \partial(|w-j|) } < 0$, which indicates that with the increase of distance between $w$ and $j$, the value of $f(w,j)$ is decreasing. Functions $f_1(w,j)$ and $f_2(w,j)$ are two candidates for $f(w,j)$:
\begin{eqnarray}
	&& f_1(w,j) = 2^{-(|w-j|+1)}\\
	&&f_2(w,j) = (2+|w-j|)^{-1}
\end{eqnarray}
\begin{figure}[h]
	\begin{center}
		\includegraphics[width=2.7in,height=2.7in,keepaspectratio]{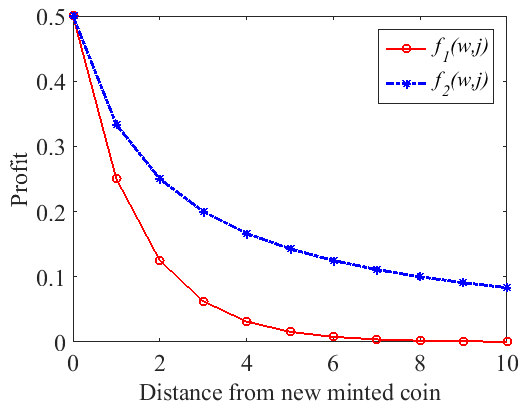} 
		\caption{Profit sharing sample functions}
		\label{fig:profit}
	\end{center}
	\vspace{-.2in}
\end{figure}
Figure~\ref{fig:profit} shows the profit sharing of a new coin using $f_1(w,j)$ and $f_2(w,j)$ candidate functions. As can be seen, with the increase of the distance of a new coin's rating and the rating of stakeholder's coin, the share of profit is decreasing. 

Let $p_{w}^{t}$ represent the profit a user earns from staking a coin $c_w$ at time $t$. Then \textit{RewardRating} calculates $p_{w}^{t}$ as follows:

\begin{equation} 
p_{w}^{t} = \frac{\gamma \times f(w,j)}{ \sum_{\forall q \in \mathcal{W}_j^t} (\sum_{\forall u_k \in \mathcal{U}} x_{k,q}^t \times f(q,j)  )  }
\label{eq:prof}
\end{equation}

To keep the system budget-balanced and simple, we assume there is a defined decimal point for the receiving profit, and the remainder of profit is received by the rating system's owner. For example, we can set 2 decimal points for the reward; in this case, if we need to divide $\$1$ to three stakeholders with the equal share, each of stakeholders receives $\$0.33$, and the rating system's owner receives $\$0.01$ as profit.  

This design provides incentives for the reviewers who correctly predict the future investments in ratings same as stock market. We analyze the benefits of such a profit sharing model in the next section. The following example is given to clarify the profit sharing scheme.

\subsection{Example}
Assume \textit{RewardRating}'s parameters have been set for a restaurant as $\alpha= 2 , \beta= 1 , \gamma= 1, n=5 $, and $f(w,j)= 2^{-(|w-j|+1)}$. 
The profit of selling the first coins are given to the rating system. Reviewers buy coins based on their prediction of future reviewers' investments. Assume at time $t$, we have $|c_1| =4 , |c_2|= 4, |c_3|= 2, |c_4|= 1$, and $|c_5|=0$. In this case, the aggregated score is:
\begin{eqnarray*}
	\sigma^t  =  \frac{ (4 \times 1) + (4 \times 2) + (3 \times 2) + (1 \times 4) + (0 \times 5) }{ 4+4+2+1 } 
\end{eqnarray*}

Assume, we set 2 decimal points for the aggregated score, then we have $\sigma^t =2.00$. The restaurant owner is malicious and wants to improve the restaurant's rating, therefore, he buys a $c_5$ coin. For the purchase of a $c_5$ coin, $1$ profit is shared among stakeholders of coins $c_3, c_4$, and $c_5$ following definition~\ref{eq:win}. Therefore, the owner of $c_4$ coin receives $0.5$, and the owner of a $c_2$ coin receives $0.25$ as reward following equation~\ref{eq:prof}. Then, the aggregated score updates to $\sigma^{t'} = 2.25$. 
Later on, an honest user predicts that the aggregated score of the service will be decreased and she buys a $c_2$ coin. At this point, the reward is shared among stakeholders of coins $c_1$, and $c_2$ following definition~\ref{eq:win}. In this case, the owner of a $c_2$ coin receives $0.16$ and the owner of a $c_1$ coin receives $0.08$ and the rating system receives the $\$0.02$ profit. Then, the aggregated score updates to $\sigma^{t''} = 2.23$. 

\section{Mechanism Analysis}\label{ana}
In this section, we analyze \textit{RewardRating}. We check budget-balanced and incentive-compatibility features. Afterward, we investigate the increase of attackers' cost. Finally, we discuss the limitations.

\vspace{.1in}
\textbf{\textit{Proposition 1.}} \textit{RewardRating} satisfies the \textit{budget-balanced} property. 

\begin{proof}
	We need to show that the total input assets into the system is equal to the total output.  
	Input assets to the system are total coins the system sells to users, and total output is the money that the system pays to the users to buy their coins in addition to the profit, which is shared among users and the reward system. For every coin, we have $\alpha = \beta + \gamma$; if we show that the total profit shared among all of the stakeholders for all of their coins is equal to $\gamma$, then we can conclude that the system is budget-balanced. Note that, the profit of selling the first coin is received by the reward system.
	For every new minted coin $c_j \in C$, the total profit which is shared among stakeholders is as follows:
	\begin{eqnarray*}
		&& \sum_{\forall w \in \mathcal{W}_j^t} (\sum_{\forall u_l \in \mathcal{U}} x_{l,w}^t \times p_w^t)= \\
		&& \sum_{\forall w \in \mathcal{W}_j^t} (\sum_{\forall u_l \in \mathcal{U}} x_{l,w}^t \times  \frac{\gamma \times f(w,j)}{ \sum_{\forall q \in \mathcal{W}_j^t} (\sum_{\forall u_k \in \mathcal{U}} x_{k,q}^t \times f(q,j)  )  }) \\
		&& =  \gamma \times   ( \frac{ \sum_{\forall w \in \mathcal{W}_j^t} ( \sum_{\forall u_l \in \mathcal{U}} x_{l,w}^t \times f(w,j))}{ \sum_{\forall q \in \mathcal{W}_j^t} (\sum_{\forall u_k \in \mathcal{U}} x_{k,q}^t \times f(q,j)  )  })   = \gamma  
	\end{eqnarray*} 
	\end{proof}
For incentive-compatibility, we need to show that \textit{RewardRating} provides incentives for honest reviewers and increases the cost for attackers. 
Same as the stock market, in \textit{RewardRating}, players' profits depend on how well they can predict the other investors' decisions in the future. This is due to the fact that the profit that a player earns from \textit{RewardRating} depends on the future investment in the rating system. 

More specifically, the \textit{RewardRating} game can be classified as follows:
\begin{itemize}
	\item \textit{Sequential}: As players choose their actions consecutively. 
	\item \textit{Perfect}: As players are aware of the previous players' actions.
	\item \textit{Non-cooperative}: As players compete to earn more profit. 
	\item \textit{Incomplete-information}: As players do not have complete knowledge about the number of players and future ratings. 
\end{itemize}
As the number of players, their types (i.e., \textit{honest}, \textit{attackers},  and \textit{strategic}), and their rating strategies are unknown, this game is classified as an \textit{Incomplete-information} game. Therefore, we analyze the strategic player's best response strategy considering different estimations for future investments.

\vspace{.1in}
\textbf{\textit{Proposition 2.}} \textit{RewardRating} satisfies the \textit{incentive-compatibility} property. 
\begin{proof}
	As there are three type of players as \textit{honest}, \textit{attackers},  and \textit{strategic} in our system model, we sketch the proof by analyzing the payoff for each type of players. 
	For honest users, \textit{RewardRating} makes profit as long as other honest users participate in the rating process. This is due to the fact that honest users are rewarded by the new honest ratings.	
	On the other hand, \textit{RewardRating} increases the cost of attack as attackers should buy coins from the system. However, attackers do not gain any profit from honest users. This is because when attackers invest in fake scores, the aggregated score is changed toward a lower or a higher score. As a result, honest reviewers will invest in the opposite direction, and attackers do not profit from such investments based on the profit-sharing model.
	Finally, for strategic users, the best response strategy is to invest in a coin in which there will be more profit in the future. Therefore, as long as the profit earned from choosing the honest rating is higher than other ratings, the best response strategy is to honestly rate the service. In this case, the incentive-compatibility requirement is met.
	
	Now, assume the strategic user estimates that the profit of investing in a fake rating is higher than an honest rating. In this case, a strategic user would invest in the attacker's side to achieve more profit. However, the profit that the system earns from selling coins to attackers will be shared among strategic users as well, following the profit-sharing model, which means that the system increases the cost for attackers. In this case, attackers endure more cost. On the other hand, this does not negatively affect honest users as they receive their profit from future honest users. Such a profit is not shared among attackers and strategic users who chose the attacker's side. Therefore, although the valid estimation of the aggregated rating is not fulfilled, the system still satisfies the incentive-compatibility feature while causing extra cost for the attacker.
	Therefore, we conclude that the proposed mechanism satisfies the incentive-compatibility feature.
\end{proof}

\subsection{Attacker's cost}
\begin{figure}[h]
	\begin{center}
		\includegraphics[width=2.7in,height=2.7in,keepaspectratio]{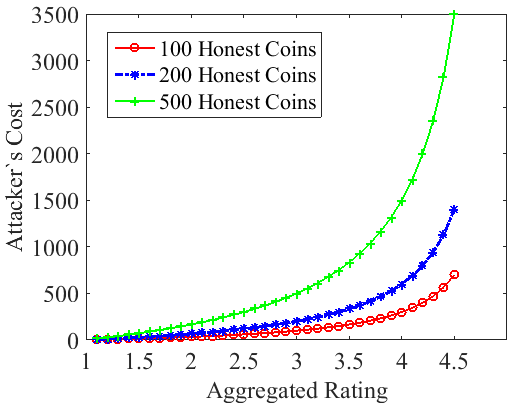} 
		\caption{Cost for attacker to increase the aggregated score of 1 with 100, 200, and 500 honest coins}
		\label{fig:cos}
	\end{center}
	\vspace{-.2in}
\end{figure}

In this section, we analyze the cost for an attacker to increase the aggregated rating. In this experiment, the initial honest aggregated score is set to 1, with three different settings of having $100$, $200$, and $500$ honest rating coins. We assume an attacker invests in the highest possible rating, which is $5$ in our example.
Figure~\ref{fig:cos} depicts the cost of an attacker for increasing the aggregated rating. As can be seen, when an attacker wants to increase the aggregated rating by investing in the highest possible rating, the cost for an attacker grows exponentially. Therefore, \textit{RewardRating}'s is more efficient for the services with a higher number of reviewers. This is due to the fact that in this case an attacker should invest in more to compensate the impact of honest ratings. 

\subsection{Limitations}
Although \textit{RewardRating} can potentially increase the cost for malicious rates while stimulating honest rates, there are two main obstacles which can preclude its adoption for micro-services.
First, \textit{RewardRating} requires reviewers to invest in ratings, and the aggregated rating is calculated based on the amount of investments in ratings. In this case, reviewers who do not want to invest in ratings cannot participate in the rating review process. 
Another concern is the rate of profit reviewers can earn from future investments. As the number of stakeholders increases, the profit earned from each new investment decreases. Thus, strategic users will not participate in the rating process when the profit they can earn is less than what they can earn in other markets. On the other hand, new honest users are reluctant to invest in ratings as the amount of profit that they can make is not decent.
One solution to overcome this problem is to set a defined value for the total profit that can be earned from staking a coin. Once this profit is achieved, the system automatically buys the coin from the corresponding stakeholder. In this case, the aggregated score should be calculated based on the number of minted coins in a specified time window as the number of coins is limited, and all of the ratings can reach out to a maximum number. By adding this feature, the total number of coins will be fixed, and as a result, the profit that a stakeholder can earn from a new investment is stabilized. Moreover, the ratings will be more dynamic as stocks are updated more quickly.

Moreover, it is worth mentioning that, although \textit{RewardRating} can increase the cost for attackers, it cannot prevent the fake ratings. Therefore, other countermeasures such as strong authentication mechanism should be applied to reduce the fake ratings. In the case that the rating system's policy limits one rating for each user, then a reviewer should not be able to purchase more than one coin from the system. 

\section{Conclusion and Future Work}\label{con}

In this paper, we have studied the challenge of designing a mechanism to preclude fake ratings while incentivizing honest ratings. First, we formally modeled the requirements for having a market for the rating system. Then, we proposed \textit{RewardRating} to increase the cost for attackers while providing incentives for honest users. 
Our analysis shows that \textit{RewardRating} satisfies budget-balanced and incentive-compatibility requirements. Our proposed mechanism can potentially increase the cost for malicious raters while stimulating honest rates. For future work, we plan to implement \textit{RewardRating} using smart-contract technology. This platform will be equipped with a browser add-on where users can use their crypto-currency assets to invest in the ratings of a service. Using smart-contract, no trusted third party is required to handle the rating scores and their corresponding assets.

%
%
%
 \bibliographystyle{splncs04}
 \bibliography{ref}

\end{document}